\documentclass{article}
\usepackage{spconf,amsmath,amssymb,bm,booktabs,cite,color,graphicx,hyperref,tabularx,url}
\hypersetup{colorlinks=true}
\makeatletter
\newcommand\footnoteref[1]{\protected@xdef\@thefnmark{\ref{#1}}\@footnotemark}
\makeatother
\title{Training Generative Adversarial Network-Based Vocoder\\
  with Limited Data Using Augmentation-Conditional Discriminator}

\name{Takuhiro Kaneko, Hirokazu Kameoka, Kou Tanaka}
\address{NTT Corporation, Japan}

\begin{document}
\ninept

\maketitle

\begin{abstract}
  A generative adversarial network (GAN)-based vocoder trained with an adversarial discriminator is commonly used for speech synthesis because of its fast, lightweight, and high-quality characteristics. However, this data-driven model requires a large amount of training data incurring high data-collection costs. This fact motivates us to train a GAN-based vocoder on limited data. A promising solution is to augment the training data to avoid overfitting. However, a standard discriminator is unconditional and insensitive to distributional changes caused by data augmentation. Thus, augmented speech (which can be extraordinary) may be considered real speech. To address this issue, we propose an \textit{augmentation-conditional discriminator (AugCond$D$)} that receives the augmentation state as input in addition to speech, thereby assessing the input speech according to the augmentation state, without inhibiting the learning of the original non-augmented distribution. Experimental results indicate that \textit{AugCond$D$} improves speech quality under limited data conditions while achieving comparable speech quality under sufficient data conditions.\footnote{\label{foot:samples}Audio samples are available at \url{https://www.kecl.ntt.co.jp/people/kaneko.takuhiro/projects/augcondd/}.}
\end{abstract}


\section{Introduction}
\label{sec:introduction}

Text-to-speech (TTS) and voice conversion (VC) have been actively studied to obtain the desired speech. In recently developed TTS and VC systems, a two-stage approach is commonly adopted, whereby the first model predicts the intermediate representation (e.g., mel spectrogram) from the input data (e.g., text or speech), and the second model synthesizes speech from the predicted intermediate representation. The second model, the neural vocoder, has been extensively studied through autoregressive models (e.g., WaveNet~\cite{AOordArXiv2016} and WaveRNN~\cite{NKalchbrennerICML2018}) and non-autoregressive models, including distillation-based (e.g., Parallel WaveNet~\cite{AOordICML2018} and ClariNet~\cite{WPingICLR2019}), flow (e.g., Glow~\cite{DKingmaNeurIPS2018})-based (e.g., WaveGlow~\cite{RPrengerICASSP2019}), diffusion~\cite{YSongNeurIPS2019,JHoNeurIPS2020}-based (e.g., WaveGrad~\cite{NChenICLR2021} and DiffWave~\cite{ZKongICLR2021}), and generative adversarial network (GAN)~\cite{IGoodfellowNIPS2014}-based (e.g., \cite{KKumarNeurIPS2019,RYamamotoICASSP2020,JKongNeurIPS2020,JYangIS2020,GYangSLT2021,AMustafaICASSP2021,JKimIS2021,TOkamotoASRU2021,TKanekoICASSP2022,MMorrisonICLR2022,TKanekoIS2022,YKoizumiSLT2022,TKanekoICASSP2023,JLeeICASSP2023,TKanekoIS2023,SDDangIS2023}) models.
This study focuses on a GAN-based model because it is fast, lightweight, and high-quality.

\begin{figure}[t]
  \centerline{\includegraphics[width=\columnwidth]{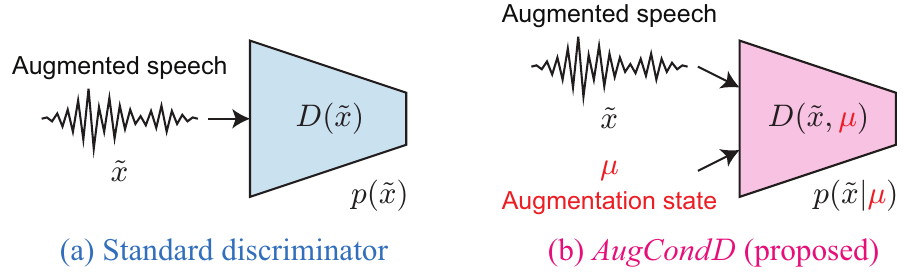}}
  \vspace{-3mm}
  \caption{Comparison of standard discriminator with proposed \textit{AugCond$D$}.
    (a)~A standard discriminator, unconditional and agnostic to the augmentation state, may consider augmented speech (which can be extraordinary) as the desired real speech.
    (b)~\textit{AugCond$D$} receives not only augmented speech but also the augmentation state, allowing it to assess the input speech conditioned on the augmentation state without interfering with the learning of the original non-augmented distribution.}
  \vspace{-4mm}
  \label{fig:teaser}
\end{figure}

A GAN-based vocoder is a data-driven model that requires a large amount of training data, resulting in high data-collection costs.
Thus, researchers have recently attempted to train the GAN-based vocoder on limited data \cite{JLeeICASSP2023}.
A promising solution is to expand the training data using data augmentation methods such as mixup~\cite{HZhangICLR2018,YTokozumeICLR2018}, CutMix~\cite{SYunICCV2019}, SpecAugment~\cite{DParkIS2019}, SpecMix~\cite{GKimIS2021}, WavAugment~\cite{EKharitonovSLT2021}, and PhaseAug~\cite{JLeeICASSP2023} to avoid overfitting.\footnote{Another possible solution is to pretrain a neural vocoder on large-scale data and then finetune it on limited target data.
  However, collecting large-scale data is often laborious and impractical in real applications as difficult ethical issues must be considered.
  Hence, in this study, we focus on training a vocoder on limited data \textit{from scratch}, deferring our ideas on finetuning to future work.}
However, as shown in Fig.~\ref{fig:teaser}(a), the standard discriminator is unconditional and agnostic to the augmentation state.
Consequently, the discriminator may consider the augmented speech (which can be extraordinary) as the desired real speech.

To address this problem, we propose an \textit{augmentation-conditional discriminator (AugCond$D$)}, which is a variant of the conditional discriminator~\cite{MMirzaArXiv2014} receiving not only speech but also the augmentation state, as shown in Fig.~\ref{fig:teaser}(b).
This allows \textit{AugCond$D$} to assess the input speech while considering the augmentation state, thereby preventing the augmented speech from interfering with the learning of the original non-augmented distribution.

In experiments, we first investigated the benchmark performance of \textit{AugCond$D$} on the LJSpeech dataset~\cite{ljspeech17}, demonstrating that \textit{AugCond$D$} improves speech quality under limited data conditions while achieving comparable speech quality under sufficient data conditions.
The general utility of \textit{AugCond$D$} was then investigated by evaluating it under various settings.

The remainder of this paper is organized as follows:
Section~\ref{sec:preliminaries} reviews a GAN-based vocoder and data augmentation.
Section~\ref{sec:augcondd} describes \textit{AugCond$D$}.
Section~\ref{sec:experiments} presents the experimental results.
Finally, Section~\ref{sec:conclusion} concludes the paper and discusses future research.

\section{Preliminaries}
\label{sec:preliminaries}

First, we review the GAN-based vocoder (Section~\ref{subsec:gan_vocoder}) and data augmentation (Section~\ref{subsec:data_augmentation}) which forms the basis of our method.

\subsection{GAN-based vocoder}
\label{subsec:gan_vocoder}

A GAN-based vocoder (or generator $G$) synthesizes speech from an intermediate representation (e.g., mel spectrogram).
It is trained with a discriminator $D$ using three losses: adversarial, feature matching, and spectrogram-domain losses.

\smallskip\noindent\textbf{Adversarial losses.}
Adversarial losses (particularly least-squares GAN-based~\cite{XMaoICCV2017} losses, which are commonly used in speech synthesis) are expressed as follows:
\begin{flalign}
  \label{eq:adv_loss_d}
  \mathcal{L}_{\mathrm{Adv}}(D) & = \mathbb{E}_{(x_r, s_r)}[ (D(x_r) - 1)^2 + (D(G(s_r)))^2 ],
  \\
  \label{eq:adv_loss_g}
  \mathcal{L}_{\mathrm{Adv}}(G) & = \mathbb{E}_{s_r}[ (D(G(s_r)) - 1)^2 ],
\end{flalign}
where $x_r$ represents real speech, and $s_r$ represents the intermediate representation (e.g., mel spectrogram) extracted from $x_r$. $D$ attempts to distinguish between real speech $x_r$ and synthesized speech $x_g = G(s_r)$ by minimizing $\mathcal{L}_{\mathrm{Adv}}(D)$.
In contrast, $G$ attempts to synthesize $x_g$ which can deceive $D$ by minimizing $\mathcal{L}_{\mathrm{Adv}}(G)$.

\smallskip\noindent\textbf{Feature matching loss.}
To stabilize GAN training, a feature matching (FM) loss~\cite{ALarsenICML2016,TKanekoIS2017b} is adopted as follows:
\begin{flalign}
  \label{eq:fm_loss}
  \mathcal{L}_{\mathrm{FM}}(G) = \mathbb{E}_{(x_r, s_r)}\left[ \sum_{i = 1}^T \frac{1}{N_i} \| D_i(x_r) - D_i(G(s_r)) \|_1 \right],
\end{flalign}
where $T$ denotes the number of layers in $D$.
$D_i$ and $N_i$ denote the features and number of features in the $i$th layer of $D$, respectively.
$G$ attempts to bring $x_g = G(s_r)$ closer to $x_r$ in the discriminator feature space by minimizing $\mathcal{L}_{\mathrm{FM}}(G)$.

\smallskip\noindent\textbf{Spectrogram-domain loss.}
A spectrogram-domain loss, such as a mel-spectrogram loss~\cite{JKongNeurIPS2020} and multiresolution spectrogram loss~\cite{RYamamotoICASSP2020}, is commonly used to further stabilize GAN training.
The mel-spectrogram loss is defined as follows:
\begin{flalign}
  \label{eq:mel_loss}
  \mathcal{L}_{\mathrm{Mel}}(G) = \mathbb{E}_{(x_r, s_r)} \left[ \| \phi(x_r) - \phi (G(s_r)) \|_1 \right],
\end{flalign}
where $\phi$ denotes a mel-spectrogram extractor.
$G$ attempts to bring $x_g = G(s_r)$ closer to $x_r$ in the mel-spectrogram domain by minimizing $\mathcal{L}_{\mathrm{Mel}}(G)$.

\smallskip\noindent\textbf{Total losses.}
The total losses are expressed as follows:
\begin{flalign}
  \label{eq:full_loss_g}
  \mathcal{L}_G & = \mathcal{L}_{\mathrm{Adv}}(G) + \lambda_{\mathrm{FM}} \mathcal{L}_{\mathrm{FM}} + \lambda_{\mathrm{Mel}} \mathcal{L}_{\mathrm{Mel}},
  \\
  \label{eq:full_loss_d}
  \mathcal{L}_D & = \mathcal{L}_{\mathrm{Adv}}(D),
\end{flalign}
where $\lambda_{\mathrm{FM}}$ and $\lambda_{\mathrm{Mel}}$ are hyperparameters for weighting the losses and were set to $2$ and $45$, respectively, in the experiments~\cite{JKongNeurIPS2020}.
$G$ and $D$ are optimized by minimizing $\mathcal{L}_G$ and $\mathcal{L}_D$, respectively.

\subsection{Data augmentation}
\label{subsec:data_augmentation}

\begin{figure}[t]
  \centerline{\includegraphics[width=\columnwidth]{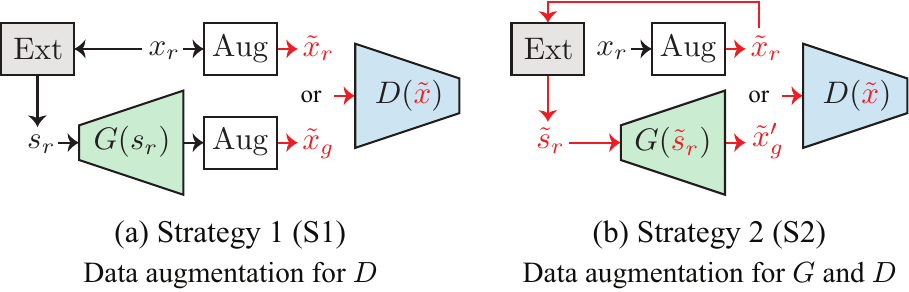}}
  \vspace{-3mm}
  \caption{Comparison of data-augmentation strategies.
    ``$\mathrm{Ext}$'' and ``$\mathrm{Aug}$'' denote an intermediate representation extractor and augmentation operator, respectively.
    The red variable and red arrow indicate augmented data and augmented data flow, respectively.
    Two data-augmentation strategies can be considered for the GAN-based vocoder: (a) augmenting only the training data for $D$; (b) augmenting data for both $G$ and $D$.}
  \label{fig:augmentation}
  \vspace{-4mm}
\end{figure}

The data augmentation technique prevents overfitting by expanding the training data.
In the context of a GAN-based vocoder, two data-augmentation strategies can be considered, as shown in Fig.~\ref{fig:augmentation}.

\smallskip\noindent\textbf{Strategy 1 (S1):~Data augmentation for $D$ (Fig.~\ref{fig:augmentation}(a)).}
Data augmentation is applied to $x_r$ and $x_g$, that is, $\tilde{x}_r = \mathrm{Aug}(x_r)$ and $\tilde{x}_g = \mathrm{Aug}(x_g)$, where ``$\mathrm{Aug}$'' denotes the augmentation operator.
Subsequently, $\tilde{x}_r$ and $\tilde{x}_g$ are fed into $D$.
This strategy is commonly used for training a GAN generating data from random noise~\cite{SZhaoNeurIPS2020,TKarrasNeurIPS2020}.

\smallskip\noindent\textbf{Strategy 2 (S2):~Data augmentation for $G$ and $D$ (Fig.~\ref{fig:augmentation}(b)).}
Data augmentation is applied to $x_r$, that is, $\tilde{x}_r = \mathrm{Aug}(x_r)$.
The intermediate representation $\tilde{s}_r$ is extracted from $\tilde{x}_r$; $G$ receives $\tilde{s}_r$ as input and synthesizes augmented speech $\tilde{x}_g'$.
Subsequently, the augmented speech, that is, $\tilde{x}_r$ or $\tilde{x}_g'$, is fed into $D$.

In S1, only the training data for $D$ are augmented (i.e., $\tilde{x}_r$ and $\tilde{x}_g$ are used), and the inputs of $G$ are not changed (i.e., $s_r$ is used).
In contrast, in S2, not only the training data for $D$ are augmented (i.e., $\tilde{x}_r$ and $\tilde{x}_g'$ are used) but also the inputs of $G$ are augmented (i.e., $\tilde{s}_r$ is used).
The preliminary experiments indicate that S2 is more effective than S1 under limited data conditions.
This is possibly because it is important to prevent not only the overfitting of $D$ but also the overfitting of $G$.
Therefore, we adopted S2 for the remainder of this study.
When S2 is employed, $\tilde{x}_r$ and $\tilde{s}_r$ are used instead of $x_r$ and $s_r$, respectively, in Eqs.~\ref{eq:adv_loss_d}--\ref{eq:full_loss_d}.

\section{Augmentation-Conditional Discriminator}
\label{sec:augcondd}

\begin{figure}[t]
  \centerline{\includegraphics[width=\columnwidth]{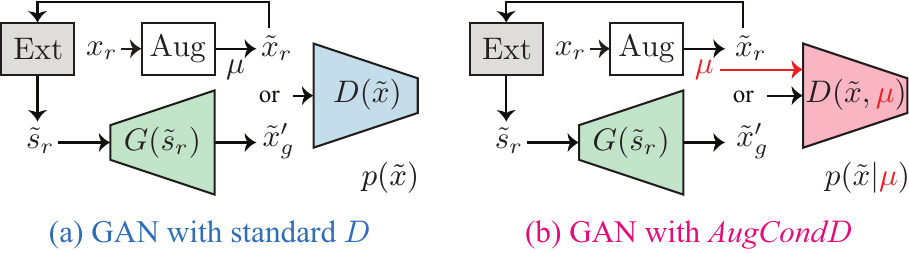}}
  \vspace{-3mm}
  \caption{Comparison of process flows for a GAN with a standard discriminator and GAN with \textit{AugCond$D$}.
    (a) Standard discriminator $D(\tilde{x})$ receives augmented speech $\tilde{x}$ only and is agnostic to the augmentation state $\mu$.
    (b) AugCond$D$ $D(\tilde{x}, \mu)$ accepts $\mu$ in addition to $\tilde{x}$, allowing
AugCond$D$ to assess $\tilde{x}$ while considering $\mu$.}
  \label{fig:gan}
  \vspace{-4mm}
\end{figure}

Data augmentation is useful for expanding training data and preventing overfitting.
However, as shown in Fig.~\ref{fig:gan}(a), a standard discriminator $D(\tilde{x})$, as defined in Section~\ref{sec:preliminaries}, is unconditional, that is, it receives augmented speech $\tilde{x}$ only and is agnostic to the augmentation state $\mu$.
Thus, it may consider the augmented speech (which can be extraordinary) $\tilde{x} \sim p(\tilde{x}) $ as real speech $x \sim p(x)$.
This is problematic when training data are limited because strong data augmentation, which is likely to result in extraordinarily augmented speech, is required to prevent overfitting.

\textit{AugCond$D$} was developed to prevent this undesirable phenomenon.
As shown in Fig.~\ref{fig:gan}(b), \textit{AugCond$D$} $D(\tilde{x}, \mu)$ receives $\tilde{x}$ and $\mu$.
This simple but critical modification allows \textit{AugCond$D$} to assess $\tilde{x}$ considering $\mu$, which is useful for preventing augmented speech $\tilde{x} \sim p(\tilde{x})$ from interfering with the learning of the original non-augmented distribution $p(x)$.

\smallskip\noindent\textbf{Losses.}
When \textit{AugCond$D$} is used, the losses defined in Section~\ref{subsec:gan_vocoder} (Eqs.~\ref{eq:adv_loss_d}--\ref{eq:mel_loss}) are rewritten as follows:
\begin{flalign}
  \label{eq:adv_loss_d_aug}
  \mathcal{L}_{\mathrm{Adv}}^{\mathrm{Aug}}(D) & = \mathbb{E}_{(\tilde{x}_r, \tilde{s}_r, \mu)}[ (D(\tilde{x}_r, \mu) - 1)^2 + (D(G(\tilde{s}_r), \mu))^2 ],
  \\
  \label{eq:adv_loss_g_aug}
  \mathcal{L}_{\mathrm{Adv}}^{\mathrm{Aug}}(G) & = \mathbb{E}_{(\tilde{s}_r, \mu)}[ (D(G(\tilde{s}_r), \mu) - 1)^2 ],
  \\
  \label{eq:fm_loss_aug}
  \mathcal{L}_{\mathrm{FM}}^{\mathrm{Aug}}(G) & = \mathbb{E}_{(\tilde{x}_r, \tilde{s}_r, \mu)}\left[ \sum_{i = 1}^T \frac{1}{N_i} \| D_i(\tilde{x}_r, \mu) - D_i(G(\tilde{s}_r), \mu) \|_1 \right],
  \\
  \label{eq:mel_loss_aug}
  \mathcal{L}_{\mathrm{Mel}}^{\mathrm{Aug}}(G) & = \mathbb{E}_{(\tilde{x}_r, \tilde{s}_r)} \left[ \| \phi(\tilde{x}_r) - \phi (G(\tilde{s}_r)) \|_1 \right].
\end{flalign}
Accordingly, the total losses (Eqs.~\ref{eq:full_loss_g} and \ref{eq:full_loss_d}) are redefined as follows:
\begin{flalign}
  \label{eq:full_loss_g_aug}
  \mathcal{L}_G^{\mathrm{Aug}} & = \mathcal{L}_{\mathrm{Adv}}^{\mathrm{Aug}}(G) + \lambda_{\mathrm{FM}} \mathcal{L}_{\mathrm{FM}}^{\mathrm{Aug}} + \lambda_{\mathrm{Mel}} \mathcal{L}_{\mathrm{Mel}}^{\mathrm{Aug}},
  \\
  \label{eq:full_loss_d_aug}
  \mathcal{L}_D^{\mathrm{Aug}} & = \mathcal{L}_{\mathrm{Adv}}^{\mathrm{Aug}}(D),
\end{flalign}
where $G$ and $D$ are optimized by minimizing $\mathcal{L}_G^{\mathrm{Aug}}$ and $\mathcal{L}_D^{\mathrm{Aug}}$, respectively.
Notably, in implementation, the required modifications are limited to changes in the inputs of $G$ and $D$ and the architectural change in $D$ (detailed below), without necessitating other modifications (e.g., changes in the weight update rule).

\smallskip\noindent\textbf{Architecture.}
In \textit{AugCond$D$}, $D$ is conditional on $\mu$ using input concatenation~\cite{MMirzaArXiv2014} because preliminary experiments indicate that this works sufficiently well. Fig.~\ref{fig:conditioning} depicts the conditioning method.
First, $\mu$ is reshaped such that its time length is identical to that of the input speech $\tilde{x}$.
Subsequently, $\mu$ and $\tilde{x}$ are concatenated in the channel direction.
Finally, the concatenated tensor is input into $D$.
The increases in calculation cost and model size are not significant because the modification is limited to changes in the input.

\section{Experiments}
\label{sec:experiments}

Two experiments were conducted to verify the effectiveness of \textit{AugCond$D$}.
(1) The benchmark performance of \textit{AugCond$D$} was investigated on the LJSpeech dataset~\cite{ljspeech17} (Section~\ref{subsec:benchmark}).
(2) The general utility of \textit{AugCond$D$} was evaluated under various settings (Section~\ref{subsec:general_utility}).
Audio samples are available from the link indicated on the first page of this manuscript.\footnoteref{foot:samples}

\subsection{Investigation of benchmark performance}
\label{subsec:benchmark}

\begin{figure}[t]
  \centerline{\includegraphics[width=\columnwidth]{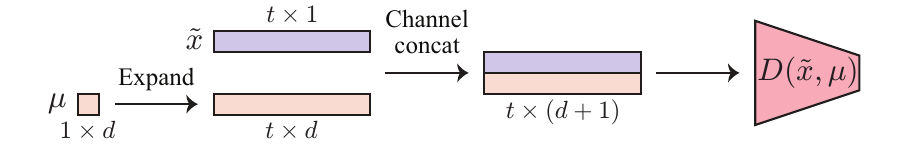}}
  \vspace{-3mm}
  \caption{Process of input concatenation.
    $A \times B$ indicates a tensor shape with time length $A$ and $B$ channels; $t$ and $d$ denote the time length of augmented speech $\tilde{x}$ and the dimension of augmentation state $\mu$, respectively.
    After $\mu$ is expanded by a factor of $t$ in the temporal direction, it is concatenated with $\tilde{x}$ in the channel direction.
    Finally, the concatenated tensor is input into $D$.
    When $\mu$ is a scalar (as in the experiments), $\mu$ is first expanded to a $1 \times 1$ tensor (i.e., $d = 1$) and then the above process is adopted.}
  \label{fig:conditioning}
  \vspace{-4mm}
\end{figure}

\textbf{Comparison models.}
In our experiments, four models were compared:
(1) HiFi-GAN (\textit{HiFi}) (particularly the high-quality V1 variant)~\cite{JKongNeurIPS2020} is a commonly used baseline.
(2) HiFi-GAN with PhaseAug (\textit{HiFi-phase})~\cite{JLeeICASSP2023} is a state-of-the-art model trained with limited data.
(3) HiFi-GAN-\textit{AugCond$D$} with mixup (\textit{HiFi-ACD-mix}) is the proposed model.
(4) HiFi-GAN with mixup (\textit{HiFi-mix}) is an ablation of (3), where a standard discriminator is used instead of \textit{AugCond$D$} to investigate the effectiveness of \textit{AugCond$D$}.
For a fair comparison of (1) and (2), (3) was implemented based on HiFi-GAN with the same generator and discriminator,  only modifying the input layer of the discriminator according to the process presented in Fig.~\ref{fig:conditioning}.
As a data augmentation method, \textit{mixup}~\cite{HZhangICLR2018} was adopted in the waveform domain because preliminary experiments indicated that it is more effective than PhaseAug (state-of-the-art) under limited-data conditions.\footnote{This is possibly because PhaseAug augments the phase of a waveform and its effect disappears when the mel spectrogram is extracted, implying that this is only effective for $D$ and not for $G$ even when S2 (Fig.~\ref{fig:augmentation}(b)) is used.
  In contrast, the mixup is effective for both $G$ and $D$ when S2 is used.}
In particular, we obtain augmented speech $\tilde{x}_r$ by mixing two speech samples in a batch ($x_r^1$ and $x_r^2$) with a mixture rate of $m \sim U(0, 1)$, that is, $\tilde{x}_r = m x_r^1 + (1 - m) x_r^2$, where $U(a, b)$ is a uniform distribution in $[a, b]$.
We defined the augmentation state $\mu$ as $\mu = (1 - \max(m, 1-m)) \times 2$.
According to this definition, $\mu = 0$ (i.e., $m = 0$ or $m = 1$) and $\mu = 1$ (i.e., $m = 0.5$) indicate no augmentation and maximum augmentation, respectively.
As $\mu$ is scalar, the process depicted in Fig.~\ref{fig:conditioning} is adopted after expanding to a $1 \times 1$ tensor (i.e., $d = 1$).

\smallskip\noindent\textbf{Data.}
The performances of the models were investigated using the LJSpeech dataset~\cite{ljspeech17}, which is a commonly used benchmark dataset.
This dataset includes 13,100 audio clips from a single female English speaker.
The clips were divided into 12,950 and 150 audio clips for training and validation, respectively, following the official HiFi-GAN~\cite{JKongNeurIPS2020} configuration.\footnote{\label{foot:hifi-gan}\url{https://github.com/jik876/hifi-gan}} 
Following the official PhaesAug~\cite{JLeeICASSP2023} configuration,\footnote{\label{foot:phaseaug}\url{https://github.com/maum-ai/phaseaug}} $100\%$ (23.7 hr.) and $1\%$ (14.4 min.) of the training data were used to simulate the sufficient and limited data conditions, respectively.
Audio clips were sampled at 22.05 kHz.
80-dimensional log-mel spectrograms extracted from audio clips with an FFT size of $1024$, hop length of $256$, and window length of $1024$ were used as vocoder inputs.

\smallskip\noindent\textbf{Implementation.}
The models were implemented using the official HiFi-GAN\footnoteref{foot:hifi-gan} and PhaseAug\footnoteref{foot:phaseaug} codes.
Each model was trained for 2.5M iterations using the Adam optimizer~\cite{DPKingmaICLR2015} with a batch size of $16$, an initial learning rate of $0.0002$, and momentum terms $\beta_1$ and $\beta_2$ of $0.5$ and $0.9$, respectively.
As reported in~\cite{JLeeICASSP2023}, \textit{HiFi} and \textit{HiFi-phase} suffered from overfitting in early iterations; therefore, we also evaluated their early stopped versions (stopped at 200k iterations).
Hereafter, these models are labeled as $\dag$.

\smallskip\noindent\textbf{Evaluation metrics.}
A mean opinion score (\textit{MOS}) test was conducted to evaluate speech quality.
Twenty audio clips were randomly selected from the validation set, and the log-mel spectrograms extracted from the audio clips were used as vocoder inputs.
In addition to the synthesized speech, \textit{ground-truth} speech was included as anchor data.
Fifteen participants attended the online test and were asked to rate speech quality on a five-point scale: 1 = bad, 2 = poor, 3 = fair, 4 = good, and 5 = excellent.
We further used three objective metrics:
(1)~\textit{UTMOS~\cite{TSaekiIS2022}} is a MOS prediction system that achieved the highest score for several metrics in the VoiceMOS Challenge 2022~\cite{WHHuangIS2022}.
Higher values correspond to better speech quality.
(2)~\textit{Periodicity~\cite{MMorrisonICLR2022}} measures the difference in periodicity between the synthesized and ground-truth speech.
This correlates with MOS~\cite{MMorrisonICLR2022} with a lower value indicating a higher degree of similarity.
(3)~Conditional Fr\'{e}chet wav2vec distance (\textit{cFW2VD})~\cite{TKanekoICASSP2022} measures the distribution distance between the synthesized and ground-truth speech in a wav2vec 2.0~\cite{ABaevskiNeurIPS2020} feature space conditioned on the text.
This correlates with MOS~\cite{TKanekoICASSP2022} such that a smaller value indicates a higher degree of similarity.

\smallskip\noindent\textbf{Results.}
Table~\ref{tab:result_ljspeech} presents the results.
As shown, the proposed model (\textit{HiFi-ACD-mix}) outperformed the other models for all metrics under limited data conditions (with a data ratio of 1\%).
\textit{AugCond$D$} achieved a performance comparable to that of the best model under sufficient data conditions (with a data ratio of 100\%).\footnote{For MOS, we conducted the Mann--Whitney U test.
  The results indicate that \textit{HiFi-ACD-mix} is significantly better than the other models under limited data conditions and is \textit{not} significantly different from the best model (i.e., \textit{HiFi-phase}) under sufficient data conditions for a $p$ value of $0.05$.}
These results indicate that the proposed method can be used for various data conditions without adverse effects.

\begin{table}[h]
  \vspace{-3mm}
  \caption{Comparison of MOS with 95\% confidence interval, UTMOS, periodicity, and cFW2VD for LJSpeech.
    In the ``MOS'' column, bold font is used when the corresponding model is \textit{not} significantly different from the best model on the Mann--Whitney U test.}
  \label{tab:result_ljspeech}
  \vspace{1mm}
  \newcommand{\spm}[1]{{\tiny$\pm$#1}}  
  \setlength{\tabcolsep}{5pt}
  \centering
  \scriptsize{
    \begin{tabularx}{\columnwidth}{lccccc}
      \toprule
      \multicolumn{1}{c}{\textbf{Model}} & \textbf{Data} & \textbf{MOS}$\uparrow$ & \textbf{UTMOS}$\uparrow$ & \textbf{Periodicity}$\downarrow$ & \textbf{cFW2VD}$\downarrow$
      \\ \bottomrule\addlinespace[\belowrulesep]
      Ground truth & --
      & 4.69\spm{0.07} & 4.38 & -- & --
      \\ \bottomrule\addlinespace[\belowrulesep]
      HiFi & 100\%
      & \textbf{4.48}\spm{0.08} & \textbf{4.23} & 0.106 & 0.022
      \\
      HiFi-phase & 100\%
      & \textbf{4.49}\spm{0.08} & \textbf{4.23} & \textbf{0.105} & 0.023    
      \\
      HiFi-mix & 100\%
      & 4.35\spm{0.09} & 4.19 & 0.108 & 0.023
      \\
      \textbf{HiFi-ACD-mix} & 100\%
      & \textbf{4.42}\spm{0.09} & \textbf{4.23} & 0.107 & \textbf{0.020}
      \\ \bottomrule\addlinespace[\belowrulesep]
      HiFi & 1\%
      & 2.89\spm{0.12} & 3.47 & 0.168 & 0.090
      \\
      HiFi$\dag$ & 1\%
      & 3.53\spm{0.12} & 3.75 & 0.143 & 0.079
      \\
      HiFi-phase & 1\%
      & 3.01\spm{0.12} & 3.46 & 0.166 & 0.091
      \\
      HiFi-phase$\dag$ & 1\%
      & 3.62\spm{0.12} & 3.71 & 0.143 & 0.073
      \\
      HiFi-mix & 1\%
      & 3.88\spm{0.11} & 3.83 & 0.125 & 0.047
      \\
      \textbf{HiFi-ACD-mix} & 1\%
      & \textbf{4.25}\spm{0.10} & \textbf{4.00} & \textbf{0.117} & \textbf{0.036}
      \\ \bottomrule
    \end{tabularx}
  }
  \vspace{-4mm}
\end{table}

\subsection{Investigation of general utility}
\label{subsec:general_utility}

The general utility of \textit{AugCond$D$} was investigated by evaluating it using different network architectures (Section~\ref{subsubsec:eval_network}), data augmentation methods (Section~\ref{subsubsec:eval_augmentation}), and speakers (Section~\ref{subsubsec:eval_speaker}).

\subsubsection{Evaluation with different network architectures}
\label{subsubsec:eval_network}

\textbf{Experimental setup.}
In Section~\ref{subsec:benchmark}, HiFi-GAN V1~\cite{JKongNeurIPS2020} was used as the baseline.
To investigate the dependence on network architecture, we evaluated \textit{AugCond$D$} using two different network architectures: HiFi-GAN V2 (\textit{HiFiV2}) (lightweight variant)~\cite{JKongNeurIPS2020} and iSTFTNet (\textit{iSTFT}) (in particular, V1-\texttt{C8C8I} (balanced variant))~\cite{TKanekoICASSP2022}.
In the experiment presented in Section~\ref{subsec:benchmark}, \textit{HiFi-mix} achieved the best performance among the baseline models under limited data conditions.
Therefore, in this experiment, we used \textit{$X$-mix} ($X$ with mixup) as the baseline, and \textit{$X$-ACD-mix} ($X$-\textit{AugCond$D$} with mixup) as an implementation of the proposed model for $X \in \{ \textit{HiFiV2}, \textit{iSTFT} \}$.
The models were trained under limited data conditions (with a data ratio of 1\%), where the training settings were identical to those presented in Section~\ref{subsec:benchmark}.

\smallskip\noindent\textbf{Results.}
Table~\ref{tab:result_network} presents the results.
The same tendencies are observed, in that, the proposed model (\textit{$X$-ACD-mix}) outperforms the baseline model (\textit{$X$-mix}) for all the metrics.
These results indicate that \textit{AugCond$D$} is not affected by network architecture.

\begin{table}[h]
  \vspace{-3mm}
  \caption{Comparison of UTMOS, periodicity, and cFW2VD for LJSpeech when HiFi-GAN V2 (\textit{HiFiV2}) and iSTFTNet (\textit{iSTFT}) are used as baselines.}
  \label{tab:result_network}
  \vspace{1mm}
  \setlength{\tabcolsep}{8pt}
  \centering
  \scriptsize{
    \begin{tabularx}{\columnwidth}{lcccc}
      \toprule
      \multicolumn{1}{c}{\textbf{Model}} & \textbf{Data} & \textbf{UTMOS}$\uparrow$ & \textbf{Periodicity}$\downarrow$ & \textbf{cFW2VD}$\downarrow$
      \\ \bottomrule\addlinespace[\belowrulesep]
      Ground-truth
      & -- & 4.38 & -- & --
      \\ \bottomrule\addlinespace[\belowrulesep]
      HiFiV2-mix
      & 1\% & 3.73 & 0.137 & 0.068
      \\
      \textbf{HiFiV2-ACD-mix}
      & 1\% & \textbf{3.81} & \textbf{0.128} & \textbf{0.052}
      \\ \bottomrule\addlinespace[\belowrulesep]
      iSTFT-mix
      & 1\% & 3.82 & 0.121 & 0.049
      \\
      \textbf{iSTFT-ACD-mix}
      & 1\% & \textbf{3.99} & \textbf{0.118} & \textbf{0.037}
      \\ \bottomrule
    \end{tabularx}
  }
  \vspace{-4mm}
\end{table}

\subsubsection{Evaluation with different data-augmentation methods}
\label{subsubsec:eval_augmentation}

\textbf{Experimental setup.}
In the aforementioned experiments, mixup~\cite{HZhangICLR2018} was used for data augmentation.
To investigate the dependence on the data augmentation method, we evaluated \textit{AugCond$D$} using another data augmentation method, \textit{speaking rate change}~\cite{EKharitonovSLT2021}, where the augmented speech $\tilde{x}_r$ is obtained by changing its speed by a factor of $2^s$ ($s \sim U(-1, 1)$).
We defined the augmentation state $\mu$ as $\mu = 2^s$, expanding it to a $1 \times 1$ tensor (i.e., $d = 1$) for the process presented in Fig.~\ref{fig:conditioning}.
HiFi-GAN with speaking rate change (\textit{HiFi-rate}) constituted the baseline, and HiFi-GAN-\textit{AugCond$D$} with speaking rate change (\textit{HiFi-ACD-rate}) was used as an implementation of the proposed model.
The models were trained under limited data conditions (with a data ratio of 1\%).
The training settings were identical to those described in Section~\ref{subsec:benchmark}.

\smallskip\noindent\textbf{Results.}
The results are presented in Table~\ref{tab:result_augmentation}.
As before, the proposed model (\textit{HiFi-ACD-rate}) outperforms the baseline (\textit{HiFi-rate}) for all metrics.
These results indicate that \textit{AugCond$D$} is effective for various data augmentation methods.

\begin{table}[h]
  \vspace{-4mm}
  \caption{Comparison of UTMOS, periodicity, and cFW2VD for LJSpeech for data augmented by \textit{speaking rate change}.}
  \label{tab:result_augmentation}
  \vspace{1mm}
  \setlength{\tabcolsep}{9pt}
  \centering
  \scriptsize{
    \begin{tabularx}{\columnwidth}{lcccc}
      \toprule
      \multicolumn{1}{c}{\textbf{Model}} & \textbf{Data} & \textbf{UTMOS}$\uparrow$ & \textbf{Periodicity}$\downarrow$ & \textbf{cFW2VD}$\downarrow$
      \\ \bottomrule\addlinespace[\belowrulesep]
      Ground-truth
      & -- & 4.38 & -- & --
      \\ \bottomrule\addlinespace[\belowrulesep]
      HiFi-rate
      & 1\% & 3.56 & 0.167 & 0.090
      \\
      \textbf{HiFi-ACD-rate}
      & 1\% & \textbf{4.10} & \textbf{0.117} & \textbf{0.033}
      \\ \bottomrule
    \end{tabularx}
  }
  \vspace{-4mm}
\end{table}

\subsubsection{Evaluation for different speakers}
\label{subsubsec:eval_speaker}

\textbf{Experimental setup.}
To investigate the speaker dependence, \textit{AugCond$D$} was evaluated using data from different speakers.
Speech of male (ID 260) and female (ID 1580) speakers was selected from LibriTTS~\cite{HZenIS2019}.
In total, 50\% of the utterances were used for the training (9.1 and 8.8 mins. for ID 260 and 1580, respectively), and the remaining data were used for validation.
\textit{HiFi-ACD-mix} and \textit{HiFi-mix} were trained individually for each speaker.
The training settings were identical to those presented in Section~\ref{subsec:benchmark} except that the training was stopped after 500k iterations because the models tended to suffer from overfitting in an earlier phase because of the small amount of training data.

\smallskip\noindent\textbf{Results.}
Table~\ref{tab:result_speaker} presents the results.
For both speakers, the proposed model (\textit{HiFi-ACD-mix}) outperforms the baseline  (\textit{HiFi-mix}) for all metrics.
These results indicate that \textit{AugCond$D$} is effective for different speakers.

\begin{table}[h]
  \vspace{-4mm}
  \caption{Comparison of UTMOS, periodicity, and cFW2VD for \textit{different speakers} in LibriTTS.}
  \label{tab:result_speaker}
  \vspace{1mm}
  \setlength{\tabcolsep}{5.5pt}
  \centering
  \scriptsize{
    \begin{tabularx}{\columnwidth}{lccccc}
      \toprule
      \multicolumn{1}{c}{\textbf{Model}} & \textbf{ID} & \textbf{Gender} & \textbf{UTMOS}$\uparrow$ & \textbf{Periodicity}$\downarrow$ & \textbf{cFW2VD}$\downarrow$
      \\ \bottomrule\addlinespace[\belowrulesep]
      Ground-truth
      & 260 & Male & 4.15 & -- & --
      \\ \bottomrule\addlinespace[\belowrulesep]
      HiFi-mix
      & 260 & Male & 3.51 & 0.150 & 0.105
      \\
      \textbf{HiFi-ACD-mix}
      & 260 & Male & \textbf{3.66} & \textbf{0.140} & \textbf{0.074}
      \\ \bottomrule \bottomrule\addlinespace[\belowrulesep]
      Ground-truth
      & 1580 & Female & 4.09 & -- & --
      \\ \bottomrule\addlinespace[\belowrulesep]
      HiFi-mix
      & 1580 & Female & 3.45 & 0.113 & 0.131
      \\
      \textbf{HiFi-ACD-mix}
      & 1580 & Female & \textbf{3.55} & \textbf{0.106} & \textbf{0.096}
      \\ \bottomrule
    \end{tabularx}
  }
  \vspace{-3mm}
\end{table}

\section{Conclusion}
\label{sec:conclusion}

We proposed \textit{AugCond$D$} to train a GAN-based vocoder with limited data.
\textit{AugCond$D$} is unique in that it receives not only speech but also the augmentation state, thereby assessing the input speech according to the augmentation state.
This prevents the augmented speech from inhibiting the learning of the original non-augmented distribution.
Experimental evaluations under various settings indicate the general utility of \textit{AugCond$D$}.
The simplicity and versatility of \textit{AugCond$D$} facilitates its application to other models (e.g., end-to-end models) and tasks (e.g., finetuning), to be pursued in future research.

\smallskip\noindent\textbf{Acknowledgements.}
This work was supported by JST CREST Grant Number JPMJCR19A3, Japan.

\vfill\pagebreak

\bibliographystyle{IEEEbib}
\renewcommand{\baselinestretch}{0.9}
{\footnotesize\bibliography{refs}}

\end{document}